\documentclass[aps,preprint]{revtex4}%
\usepackage{amsfonts}
\usepackage{amsmath}
\usepackage{amssymb}
\usepackage{graphicx}%
\providecommand{\U}[1]{\protect\rule{.1in}{.1in}}

\begin{document}
\title[ ]{Criterion for the Thermal Radiation Spectrum in Classical Physics}
\author{Timothy H. Boyer}
\affiliation{Department of Physics, City College of the City University of New York, New
York, New York 10031}

\begin{abstract}
Two criteria for the spectra of relativistic waves are proposed. Zero-point
radiation provides the identity representation of the conformal group in
Minkowski spacetime. \ Thermal radiation provides the irreducible
representation of the conformal group in Minkowski spacetime which involves
exactly one scaling parameter (the temperature) which is also time-stationary
in a Rindler frame. \ Zero-point radiation is the limit of thermal radiation
as the temperature goes to zero. \ Crucially, both zero-point radiation and
thermal radiation take basically the same functional form in a Rindler frame.
\ For relativistic scalar waves, a full derivation of the Planck spectrum
including zero-point radiation is obtained within the classical theory.

\end{abstract}
\email{tboyer@ccny.cuny.edu}
\maketitle

\section{Introduction}

Thermal radiation involves a finite amount of thermal energy spread across the
volume of the containing box. \ For \textit{relativisitic} thermal radiation
(waves where the mass vanishes), there are an infinite number of normal modes
within the containing box, and hence there must be at least two parameters for
the spectrum. \ Roughly, one parameter gives the energy per normal mode at low
frequencies and the other gives the energy per normal mode at high
frequencies. \ Within classical physics, there seems to be no criterion which
distinguishes thermal radiation from other random spectra of relativistic
radiation. \ Ever since the 19th century, physicists have searched for a
simple criterion which would distinguish the thermal radiation spectrum. \ 

In this article, we suggest that \textit{random relativistic thermal radiation
in Minkowski spacetime is the covariant representation of the conformal group
which involves exactly one variable parameter, the temperature }$T$,\textit{
and which is time-stationary in both an inertial frame and in a Rindler frame.
\ }

In an earlier article,\cite{B2003c} it was suggested that the thermal
radiation spectrum was the \textquotedblleft smoothest\textquotedblright%
\ function joining zero-point radiation in the limit of small temperature and
the Rayleigh-Jeans spectrum at high temperature. \ However, the idea of
\textquotedblleft smoothness\textquotedblright\ seems vague compared with
explicit conformal requirements. \ Any distribution of \textit{relativistic
radiation} is conformally covariant in Minkowski spacetime. \ However, the
requirement on the number of variable parameters in the irreducible
representation restricts the spectrum sharply. \ The other important aspect is
the use of a Rindler frame. \ A Rindler frame separates the time behavior from
the spatial behavior, and so provides a new perspective on relativistic
thermal radiation. A Rindler frame is special in that the spacetime intervals
though first order in the Rindler frame agree with those in\ the momentarily
comoving inertial frame. \ Indeed, although we often think of ourselves as
living in a local inertial frame, we might better regard ourselves as living
in a local Rindler frame. \ 

In order to account for Casimir forces within classical theory, we must have
zero-point radiation with a constant determined to fit the measured forces.
\ However, in a Rindler frame, any spectrum having the same functional form as
zero-point radiation will be time stationary. \ By reajusting the Rindler
coordinate time while maintaining the experimental zero-point spectrum, we are
able to determine the spectrum of thermal radiation in a Minkowski inertial
frame. \ 

In this article, we do not consider the mechanism by which equilibrium is
achieved, only the criterion for equilibrium. \ However, when a charged
particle in a Coulomb potential moves in a circuit, it must accelerate and so
loses energy by radiation emission, while it gains energy by repeated resonant
motion through classical zero-point radiation. \ When the orbiting charged
particle comes to an average situation of energy balance, then we have an
equilibrium situation.\cite{B1975}

In this article, we first point out the fundamental constants of
electromagnetism and the basis for dilation covariance. \ Then we discuss the
idea of classical electromagnetic zero-point radiation and also thermal
radiation. \ Finally, we turn to the idea of using a Rindler coordinate frame
and show that our criterion leads to the Planck spectrum within classical
physics. \ 

\section{Units and Conformal Symmetry}

A \textit{relativisitic} classical theory, such as classical electromagnetism,
can incorporate the three fundamental electromagnetic constants found in
nature: the elementary charge $e$, the speed of light in vacuum $c$, and the
scale of classical zero-point radiation $\hbar$. \ The choice of units is
determined by the choice of the unit mass $M$ giving\ unit length,
$l=e^{2}/\left(  Mc^{2}\right)  ,$ unit time, $t=e^{2}/\left(  mc^{3}\right)
,$ and unit energy as $U=Mc^{2}$. \ The constant $\hbar$ is the scale factor
for classical zero-point radiation. \ The one constant which is independent of
the choice of unit mass is the fine structure constant $e^{2}/\left(  \hbar
c\right)  .$ If one \textit{changes} the unit mass from $M$ over to
$M^{\prime}=M/\sigma$, then all lengths transform to $l^{\prime}=\sigma l$,
all times change to $t^{\prime}=\sigma t$, and all energies change to
$U^{\prime}=U/\sigma$. \ We say that the system undergoes a $\sigma_{ltU^{-1}%
}$-dilation or rescaling. \ However, the fine structure constant is unchanged
under the rescaling. \ Conformal symmetry is the group of transformations
which includes those of the Poincar$\acute{e}$ group together with the
continuous, local changes of scale corresponding to altering the choice of
unit mass $M$.\cite{Kastrup}\cite{Zee}

\section{Random Relativistic Scalar Field in an Inertial Frame \ }

\subsection{Hamiltonian of a Relativistic Scalar Field}

In this article, we consider, not the \textit{vector} electromagnetic field,
but rather the still simpler relativistic \textit{scalar} field $\phi\left(
ct,x,y,z\right)  .$ \ The electromagnetic field involves the additional
complications of two directions of polarization for a transverse
electromagnetic wave. \ Therefore a scalar field is often treated as a simpler
introductory calculation. \ This situation is seen explicitly when calculating
the conformal invariance of classical zero-point radiation.\cite{B1989} \ 

The behavior of a scalar field in four-dimensional Minkowski spacetime is
given by the Hamiltonian%
\begin{equation}
H=\frac{1}{2}\int d^{3}r\left[  \left(  \frac{\partial\phi}{\partial\left(
ct\right)  }\right)  ^{2}+\left(  \nabla\phi\right)  ^{2}\right]  ,
\label{Ham}%
\end{equation}
leading to the scalar wave equation
\begin{equation}
\frac{\partial^{2}\phi}{\partial\left(  ct\right)  ^{2}}-\nabla^{2}\phi=0.
\label{sweq}%
\end{equation}
This equation (\ref{sweq}) has the isotropic solution%
\begin{equation}
\phi\left(  ct,x,y,z\right)  =\int d^{3}k\,A(k)\cos\left[  \mathbf{k\cdot
r}-kct+\theta\left(  \mathbf{k}\right)  \right]  \label{phictxyz}%
\end{equation}
with $k=\left\vert \mathbf{k}\right\vert $. \ Since we are interested in
\textit{random} radiation, we will take the phase $\theta\left(
\mathbf{k}\right)  $ as randomly distributed on the interval $[0,2\pi)$, and
independently distributed for each normal mode specified by $\mathbf{k}$. \ 

\subsection{Energy $U\left(  ck\right)  $ of a Radiation Normal Mode}

If we consider a single radiation normal mode of frequency $\omega
_{\mathbf{n}}=ck_{\mathbf{n}}$ in a cubical box of side $L,$ then, from
equation (\ref{Ham}), we have energy
\begin{align}
U\left(  ck_{\mathbf{n}}\right)   &  =\frac{1}{2}\int d^{3}r\left[  A\left(
k_{\mathbf{n}}\right)  \right]  ^{2}2k_{\mathbf{n}}^{2}\sin^{2}\left[
\mathbf{k\cdot r}-kct+\theta\left(  \mathbf{k}\right)  \right] \nonumber\\
&  =\frac{1}{2}\int d^{3}r\left[  A\left(  k_{\mathbf{n}}\right)  \right]
^{2}k_{\mathbf{n}}^{2}=\frac{1}{2}L^{3}k_{\mathbf{n}}^{2}\left[  A\left(
k_{\mathbf{n}}\right)  \right]  ^{2} \label{Uzp}%
\end{align}

\subsection{Dilation-Invariant Spectrum}

In order to be $\sigma_{ltU^{-1}}$-dilation invariant, we expect the spectrum
to be%

\begin{equation}
U\left(  \omega\right)  =C\omega, \label{const}%
\end{equation}
where $C$ is a constant, since both $U\left(  ck_{\mathbf{n}}\right)  $ and
$\omega_{\mathbf{n}}=ck_{\mathbf{n}}$ have the same behavior under the
dilation, $U^{\prime}=U/\sigma,~\omega^{\prime}=\omega/\sigma$. \ The equation
takes the same form in any inertial frame. \ \ This relationship (\ref{const})
between energy and frequency is also both Lorentz invariant and conformal
invariant.\cite{Marshall1965}\cite{B1989}\ 

\subsection{Scale from Casimir Forces}

The expression in Eq. (\ref{const}) still involves one unknown constant $C$.
\ In the case of electromagnetic fields $\mathbf{E}$ and $\mathbf{B,}$ the
random classical zero-point radiation leads to Casimir forces between
conducting parallel plates which have been measured
experimentally.\cite{Casimir}\cite{CForces}\cite{CF2}\cite{CF3}\cite{CF4}%
\cite{CF5}\cite{CF6} \ The experimental results fit (to an accuracy of close
to 1\% for both the separation between the conducting plates and also the
magnitude) with the choice for the electromagnetic energy per normal mode
\begin{equation}
U^{elect}\left(  \omega\right)  =\left(  \hbar/2\right)  \omega, \label{hbar}%
\end{equation}
where $\hbar$ is a numerical constant which agrees in numerical value with
Planck's constant. \ The constant $\hbar$ appeared first, not quantum theory,
but in the classical analysis of blackbody radiation in 1899.\cite{constant}
\ The \textit{classical} electromagnetic scale factor $\hbar$ in Eq.
(\ref{hbar}) has nothing to do with energy quanta. \ However, some physicists
have never seen $\hbar$ in any context other than quantum theory, and so are
offended to find it a \textit{classical} theory. \ 

In order to make the natural connection between the electromagnetic field and
the \textit{scalar} field case for the correlation functions, we will assume
that equation (\ref{phictxyz}) can be rewritten for scalar zero-point
radiation as%
\begin{equation}
\phi^{zp}\left(  ct,x,y,z\right)  =\int d^{3}k\,\sqrt{\frac{\hbar c}{2\pi
^{2}k}}\cos\left[  \mathbf{k\cdot r}-kct+\theta\left(  \mathbf{k}\right)
\right]  .
\end{equation}

\section{The Zero-Point Correlation Function}

\subsection{The Two-Point Field Correlation Function}

The two-point correlation function for a scalar zero-point radiation is given
by%
\begin{align}
&  \left\langle \phi^{zp}\left(  ct,x,y,z\right)  \phi^{zp}\left(  ct^{\prime
},x^{\prime},y^{\prime},z^{\prime}\right)  \right\rangle _{\theta}\nonumber\\
&  =\left\langle \int d^{3}k\,\sqrt{\frac{\hbar c}{2\pi^{2}k}}\cos\left[
\mathbf{k\cdot r}-kct+\theta\left(  \mathbf{k}\right)  \right]  \int
d^{3}k^{\prime}\,\sqrt{\frac{\hbar c}{2\pi^{2}k^{\prime}}}\cos\left[
\mathbf{k}^{\prime}\mathbf{\cdot r}^{\prime}-k^{\prime}ct^{\prime}%
+\theta\left(  \mathbf{k}^{\prime}\right)  \right]  \right\rangle _{\theta
}\nonumber\\
&  =\int d^{3}k\,\sqrt{\frac{\hbar c}{2\pi^{2}k}}\int d^{3}k^{\prime}%
\,\sqrt{\frac{\hbar c}{2\pi^{2}k^{\prime}}}\left\langle \cos\left[
\mathbf{k\cdot r}-kct+\theta\left(  \mathbf{k}\right)  \right]  \cos\left[
\mathbf{k}^{\prime}\mathbf{\cdot r}^{\prime}-k^{\prime}ct^{\prime}%
+\theta\left(  \mathbf{k}^{\prime}\right)  \right]  \right\rangle _{\theta
}\nonumber\\
&  =\int d^{3}k\sqrt{\frac{\hbar c}{2\pi^{2}k}}\int d^{3}k^{\prime}%
\,\sqrt{\frac{\hbar c}{2\pi^{2}k^{\prime}}}\cos\left[  \mathbf{k\cdot}\left(
\mathbf{r-r}^{\prime}\right)  -k\left(  ct-ct^{\prime}\right)  \right]
\frac{1}{2}\delta^{3}\left(  \mathbf{k-k}^{\prime}\right) \nonumber\\
&  =\frac{1}{2}\int d^{3}k\,\frac{\hbar c}{2\pi^{2}k}\cos\left[
\mathbf{k\cdot}\left(  \mathbf{r-r}^{\prime}\right)  -k\left(  ct-ct^{\prime
}\right)  \right]  , \label{corrtx}%
\end{align}
where we have summed over the random phases as%
\begin{equation}
\left\langle \cos\left[  \theta\left(  \mathbf{k}\right)  \right]  \cos\left[
\theta\left(  \mathbf{k}^{\prime}\right)  \right]  \right\rangle =\left\langle
\sin\left[  \theta\left(  \mathbf{k}\right)  \right]  \sin\left[
\theta\left(  \mathbf{k}^{\prime}\right)  \right]  \right\rangle =\left(
1/2\right)  \delta^{3}(\mathbf{k-k}^{\prime})
\end{equation}
and
\begin{equation}
\left\langle \cos\left[  \theta\left(  \mathbf{k}\right)  \right]  \sin\left[
\theta\left(  \mathbf{k}^{\prime}\right)  \right]  \right\rangle =0.
\end{equation}

Now we introduce $\cos\psi$ as the angle between $\mathbf{k}$ and $\left(
\mathbf{r-r}^{\prime}\right)  $ and integrate over all angles to obtain from
Eq. (\ref{corrtx})
\begin{align}
&  \left\langle \phi^{zp}\left(  ct,x,y,z\right)  \phi^{zp}\left(  ct^{\prime
},x^{\prime},y^{\prime},z^{\prime}\right)  \right\rangle \nonumber\\
&  =\frac{\hbar c}{4\pi^{2}}\int_{0}^{\infty}dk\,k\left(  2\pi\right)
\int_{0}^{\pi}d\psi\sin\psi\cos\left[  k\left\vert \mathbf{r-r}^{\prime
}\right\vert \cos\psi-k\left(  ct-ct^{\prime}\right)  \right] \nonumber\\
&  =\frac{\hbar c}{2\pi\left\vert \mathbf{r-r}^{\prime}\right\vert }\int%
_{0}^{\infty}dk\left\{  \sin\left[  k\left\{  \left\vert \mathbf{r-r}^{\prime
}\right\vert -\left(  ct-ct^{\prime}\right)  \right\}  \right]  +\sin\left[
k\left\{  \left\vert \mathbf{r-r}^{\prime}\right\vert +\left(  ct-ct^{\prime
}\right)  \right\}  \right]  \right\}  \label{corrtx2}%
\end{align}
The last line in Eq. (\ref{corrtx2}) involves singular integrals which we
evaluate by introducing a small convergence factor so that we have%

\begin{equation}
\int_{0}^{\infty}dk\sin\left[  kb\right]  \exp\left[  -ik\epsilon\right]
=\left(  \frac{-\cos\left[  k\left(  b-i\epsilon\right)  \right]
}{b-i\epsilon}\right)  _{0}^{\infty}\rightarrow\frac{1}{b}.
\end{equation}
But then the correlation function for zero-point radiation becomes%
\begin{align}
&  \left\langle \phi^{zp}\left(  ct,x,y,z\right)  \phi^{zp}\left(  ct^{\prime
},x^{\prime},y^{\prime},z^{\prime}\right)  \right\rangle \nonumber\\
&  =\frac{\hbar c}{2\pi\left\vert \mathbf{r-r}^{\prime}\right\vert }\left(
\frac{1}{\left\{  \left\vert \mathbf{r-r}^{\prime}\right\vert -\left(
ct-ct^{\prime}\right)  \right\}  }+\frac{1}{\left\{  \left\vert \mathbf{r-r}%
^{\prime}\right\vert +\left(  ct-ct^{\prime}\right)  \right\}  }\right)
\nonumber\\
&  =\frac{-\hbar c}{\pi}\left(  \frac{1}{\left(  ct-ct^{\prime}\right)
^{2}-\left\vert \mathbf{r-r}^{\prime}\right\vert ^{2}}\right)  .
\label{corrtx3}%
\end{align}

\subsection{Conformal \textit{In}variance of Classical Zero-Point Radiation}

The correlation function in Eq. (\ref{corrtx3}) for the zero-point fields is
clearly Lorentz invariant, varying as $1/\sigma^{2}$ under dilation. \ It is
also conformal invariant. \ The scalar field $\phi^{zp}\left(
ct,x,y,z\right)  $ is conformal invariant in Minkowski spacetime. \ Indeed,
all of classical electrodynamics is conformal invariant in an inertial frame.
\ Maxwell's equations are conformal covariant and so are their
solutions.\cite{C-B} \ \ Similarly, the scalar wave equation (\ref{sweq}) in
Minkowski spacetime is conformal invariant as are the solutions.\cite{B1989}
\ In this article, we will consider uniform $\sigma_{ltU^{-1}}$-dilations of
the conformal group. \ The homogeneous wave equation (\ref{sweq}) has exactly
one invariant parameter, the speed $c$\ of the relativistic waves. \ \ 

\subsection{Conformal \textit{Co}variance for Wien's Displacement Theorem}

We now turn from the conformal \textit{invariance} of zero-point radiation to
the conformal \textit{covariance} of thermal radiation in an inertial frame.
\ The work-energy ideas of thermodynamics\cite{B2003c} are sufficient to lead
to Wien's displacement theorem involving one unknown function of temperature
$T$ and frequency $\omega=c\left\vert k\right\vert =ck,$%
\begin{equation}
U^{total}\left(  ck\right)  =Tf\left(  \frac{ck}{T}\right)  .
\end{equation}
Here $U^{total}\left(  ck\right)  $ is the average energy per normal mode at
frequency $ck$, and $f\left(  ck/T\right)  $ is some unknown function.\ \ The
$\sigma_{ltU^{-1}}$-dilation behavior is correct, since the energy $U$, the
temperature $T$, and the frequency $ck$ all transform in the same way. \ In
the limit of ever-smaller temperature compared to a fixed frequency $ck$, we
expect that the spectrum becomes the zero-point radiation spectrum,
\begin{equation}
U^{total}\left(  ck\right)  =Tf\left(  \frac{ck}{T}\right)  \rightarrow
\frac{1}{2}\hbar ck\text{ \ for }T\rightarrow0,
\end{equation}
as appearing in Eq. (\ref{hbar}).

\section{Criteria for Zero-Point Radiation and Thermal Radiation}

There is a clear criterion for \textit{zero-point} radiation in classical
physics: \textit{zero-point radiation is the relativistic random classical
radiation spectrum which is conformal invariant in an inertial frame}%
.\cite{B1989} \ The scale constant of $\hbar$ is determined by experiment.

The criterion for \textit{thermal} radiation is not nearly so clear; indeed,
it seems unrecognized in physics. \ Our suggestion is that \textit{thermal
radiation is time-stationary in a closed box in both an inertial frame and in
a Rindler frame, and is conformal covariant with the one scaling parameter,
the temperature.}

\section{Rindler Frame and Thermal Radiation}

\subsection{Rindler Frame Separates Space and Time}

In Minkowski spacetime, space and time are tied together by the speed of light
in vacuum $c$. \ However, there is another spacetime where the time component
is separated from the spatial components. \ In a Rindler frame, the spatial
separations are not changing in time, yet the time aspect is different from
the spatial aspect. \ The situation is analogous to changing to polar
coordinates and keeping the angles fixed while noting that length is
associated exclusively with the radial component. \ 

We consider a Rindler frame $\left(  \eta,\xi,y,z\right)  $ related to
Minkowski space time $\left(  ct,x,y,z\right)  $ by\cite{Schutz}%
\begin{equation}
ct=\xi\sinh\eta,\text{ \ \ }x=\xi\cosh\eta,~~y=y,~~z=z. \label{Rindl}%
\end{equation}
Notice that both the time $ct$ and the spatial coordinates $x,y,z$ in the
inertial frame will transform as a spatial coordinates $\xi,y,z$ under a
$\sigma_{ltU^{-1}}$-dilation in the Rindler frame. \ The Rindler time $\eta$
is completely independent. \ This situation corresponds to the great
difference between an inertial frame and a Rindler frame. \ On the surface of
the earth, we often think of ourselves as living in a local inertial frame.
\ It might be better to consider ourselves as living in a local Rindler frame.
\ If one drops something out the window, it accelerates. \ Also, our GPS
systems must correct for the gravitational effects on time intervals. \ 

Each \textit{fixed} spatial coordinate $\xi$ of the Rindler frame has the same
speed relative to the momentarily comoving inertial frame
\begin{equation}
\frac{dx}{d\left(  ct\right)  }=\frac{\xi\sinh\eta}{\xi\cosh\eta}=\tanh\eta.
\end{equation}
However, each fixed spatial coordinate $\xi$ of the Rindler frame has a
constant acceleration which is
\begin{equation}
a\left(  \xi\right)  =c^{2}/\xi,
\end{equation}
and depends on the distance from an event horizon located at $\xi=0.$ \ At a
single coordinate time $\eta$ in the Rindler frame, the Rindler frame agrees
with the comoving inertial frame's time interval and spatial separation
through \textit{first-order} in the separations. \ The varying acceleration
$a\left(  \xi\right)  =c^{2}/\xi$ with distance from the event horizon is
\textit{second-order} in time intervals and is required by relativity.

\subsection{Contrasting Aspects of Zero-Point and Thermal Radiation}

A Rindler frame gives one a new perspective on thermal radiation. \ Zero-point
radiation and thermal radiation seem very different. \ Zero-point radiation
involves a \textit{divergent} energy density which has the same correlation
function in \textit{every} inertial frame. \ Thermal radiation has a
\textit{finite} energy density which is in equilibrium in \textit{exactly one}
inertial frame, that of the confining box where it came to equilibrium. \ At
any frequency $\omega=ck,$ total thermal radiation involves both
contributions
\begin{equation}
U^{total}\left(  \omega,T\right)  =U^{T}\left(  \omega,T\right)
+U^{zp}\left(  \omega\right)  .
\end{equation}
As emphasized later, in any box in a Rindler frame, the temperature of the
thermal radiation at the \textit{top} of the box is \textit{smaller} than the
temperature of the thermal radiation at the bottom of the box, but the
zero-point radiation radiation is the same. \ 

\section{Different Normal Modes in an Inertial Frame and a Rindler Frame}

Now if the random radiation in a closed box with perfectly conducting walls is
written in terms of radiation normal modes with random phases, then the
correlation function will not change in time. \ However, an inertial frame and
a Rindler frame do \textit{not} have the same normal modes for radiation, and,
accordingly, the correlation functions seem quite different.

\subsection{Radiation Modes Propagating Normal to the Event Horizon}

\subsubsection{Radiation Modes in an Inertial Frame}

We can see the contrasting situations easily for a special case. \ Consider a
scalar plane wave propagating exactly normal to the event horizon. \ In an
inertial frame, the plane wave is%
\begin{equation}
\phi(ct,x,y,z)=A\left(  k\right)  \cos\left[  \pm kx-ckt+\theta\left(  \pm
k\right)  \right]  . \label{unprim}%
\end{equation}
corresponding to a normal mode with frequency $\omega=ck.$ \ Under a
$\sigma_{ltU^{-1}}$-dilation, this wave becomes a new wave $\phi(ct^{\prime
},x^{\prime},y^{\prime},z^{\prime})$
\begin{equation}
\phi(ct^{\prime},x^{\prime},y^{\prime},z^{\prime})=A\left(  k\right)
\cos\left[  \pm k\left(  \sigma x\right)  -ck\left(  \sigma t\right)
+\theta\left(  \pm k\right)  \right]  . \label{phiprime}%
\end{equation}
Note the $\sigma x$ and $\sigma t$, while the phase $\theta\left(  \pm
k\right)  $ is unchange. \ This equation is \textit{not} the same as the old
plane wave in Eq. (\ref{unprim}). \ In Rindler coordinates, this plane wave
(\ref{unprim}) becomes
\begin{align}
\phi\left(  \xi\sinh\eta,\xi\cosh\eta,y,z\right)   &  =A\left(  k\right)
\cos\left[  \pm k\xi\cosh\eta-k\xi\sinh\eta\right]  +\theta\left(  k\right)
\nonumber\\
&  =A\left(  k\right)  \cos\left[  k\xi\exp\left(  \mp\eta\right)
+\theta\left(  k\right)  \right]  .
\end{align}
This radiation field is \textit{not} a normal mode in the Rindler frame. \ 

\subsubsection{Radiation Modes in a Rindler Frame}

The scalar \textit{wave equation} (\ref{sweq}) when written in Rindler
coordinates is
\begin{equation}
\frac{\partial^{2}\phi}{\partial\xi^{2}}+\frac{1}{\xi}\frac{\partial\phi
}{\partial\xi}+\frac{\partial^{2}\phi}{\partial y^{2}}+\frac{\partial^{2}\phi
}{\partial z^{2}}-\frac{1}{\xi^{2}}\frac{\partial^{2}\phi}{\partial\eta^{2}%
}=0.
\end{equation}
For a normal mode in the Rindler frame which is propagating normal to the
event horizon in a closed box with ends at $\xi_{\mathfrak{A}}=\xi_{0}$ and at
$\xi_{\mathfrak{B}}=\xi_{0}+l,$ we find
\begin{align}
&  \phi_{n}(\eta,\xi,y,z)\\
&  =\sqrt{\frac{2}{\ln\left[  \left(  \xi_{0}+l\right)  /\xi_{0}\right]  }%
}\sin\left[  \frac{n\pi}{\ln\left[  \left(  \xi_{0}+l\right)  /\xi_{0}\right]
}\ln\left(  \frac{\xi}{\xi_{0}}\right)  \right]  \cos\left[  \frac{n\pi}%
{\ln\left[  \left(  \xi_{0}+l\right)  /\xi_{0}\right]  }\eta+\theta
_{n}\right]  .
\end{align}
Clearly, the wave vanishes at the top and bottom planes since ln$\left(
1\right)  =0\,\ $and sin$\left(  n\pi\right)  =0$ for integer $n$. \ If the
unit of length is changed as $l\rightarrow l^{\prime}=\sigma l$, corresponding
to the $\sigma_{ltU^{-1}}$-dilation, then $\xi,~\xi_{0},$ and $l$ will all
change by the same factor so that the normal mode is \textit{unchanged}. \ The
normal modes of the inertial frame and of the Rindler frame are quite
different from each other. \ 

\subsection{Correlation Function in the Rindler Frame}

The time coordinate in each spacetime frame is connected to the frequency
$\omega=c\left\vert \mathbf{k}\right\vert $ of the relativistic waves in that
frame, and so is different between an inertial frame and a Rindler frame,
since a normal mode in one frame is not a normal mode in the other. \ Thus,
the random radiation seen in the Rindler frame may have a different spectrum
from that found in the Minkowski frame. \ In particular, the acceleration of
each fixed spatial coordinate $\xi$\ will lead to a different correlation
function in the Rindler frame.

Since the zero-point function $\phi^{zp}(ct,x,y,z)$ is a \textit{scalar}
field, the correlation function in the Rindler frame for fixed spatial
coordinates $\xi,y,z$ but different times $\eta$ and $\eta^{\prime}$ is, from
Eqs. (\ref{Rindl}) and (\ref{corrtx3}),%
\begin{align}
&  \left\langle \phi^{zp}\left(  ct,x,y,z\right)  \phi^{zp}\left(  ct^{\prime
},x,y,z\right)  \right\rangle \nonumber\\
&  =\left\langle \phi^{zp}\left(  \xi\sinh\eta,\xi\cosh\eta,y,z\right)
\phi^{zp}\left(  \xi\sinh\eta^{\prime},\xi\cosh\eta^{\prime},y,z\right)
\right\rangle \nonumber\\
&  =\frac{\hbar c}{\pi}\frac{1}{\xi^{2}\left[  \left(  \cosh\eta-\cosh
\eta^{\prime}\right)  ^{2}-\left(  \sinh\eta-\sinh\eta^{\prime}\right)
^{2}\right]  }\nonumber\\
&  =\frac{\hbar c}{\pi}\frac{-1}{4\xi^{2}\sinh^{2}\left[  \left(  \eta
-\eta^{\prime}\right)  /2\right]  } \label{corrzp}%
\end{align}
The correlation function here depends only on the coordinate time difference
$\eta-\eta^{\prime}$ in the Rindler frame, and so is time stationary. \ Also,
in the Rindler frame, the constant out in front $\left(  \hbar c/\pi\right)  $
could take \textit{any positive value}, just like the constant $C$ in Eq.
(\ref{const}), and would still give a \textit{time-independent} spectrum.
\ This flexibility of the time-invariance is what allows the Planck thermal
radiation spectrum. \ 

\subsection{Zero-Point Radiation is Dilation-Invariant in Both Frames}

This expression (\ref{corrzp}) for zero-point radiation in a Rindler frame is
\textit{different} from that in a Minkowski inertial frame given in Eq.
(\ref{corrtx3}). \ However, the zero-point correlation function agrees between
the Rindler frame and the comoving inertial frame, since the zero-point
radiation spectrum is Lorentz invariant in Minkowski spacetime. \ Thus,
\textit{any} inertial frame will find the same correlation function for the
zero-point radiation field. \ At small time differences $\eta-\eta^{\prime},$
the correlation function (\ref{corrzp}) in the Rindler frame goes over to the
correlation function (\ref{corrtx3}) for the inertial frame; from Eq.
(\ref{Rindl}), we see to first order in the differences $ct=\xi\sinh
\eta\approxeq\xi\eta$ and $x=\xi\cosh\eta\approxeq\xi$, so that
\begin{equation}
\frac{-\hbar c}{\pi}\frac{1}{\left(  ct-ct^{\prime}\right)  ^{2}}%
\approxeq\frac{-\hbar c}{\pi}\frac{1}{\xi^{2}\left(  \eta-\eta^{\prime
}\right)  ^{2}}. \label{limit}%
\end{equation}
Notice that for coincidence of the two spacetime \textit{points}, $t^{\prime
}\rightarrow t\,\ $\ and $\eta^{\prime}\rightarrow\eta.$ both correlation
functions diverge corresponding to the divergence of zero-point energy.
\ Also, in both an inertial frame and a Rindler frame, the correlation
functions change as $1/\sigma^{2}$ under a $\sigma_{ltU^{-1}}$-dilation. \ 

\section{Random Waves in a Box}

\subsection{Spectrum Remains Zero-Point Radiation in the Comoving Inertial
Frame}

The situation involving zero-point radiation has already been considered
several times with varying points of view in classical physics.\cite{B2002d}%
\cite{B42}\cite{Bacc} \ We emphasize that at a single time $t^{\prime}$ in the
comoving inertial frame, the spectrum of \textit{zero-point} radiation has no
preferred length or energy. \ \ 

\subsection{Thermal Radiation in a Rindler Frame}

\subsubsection{Thermal Radiation Above Zero-Point Radiation}

Classical zero-point radiation does not simply disappear when we discuss
thermal radiation. \ Rather we expect that classical thermal radiation
$U^{T}\left(  \omega,T\right)  $ is radiation \textit{in addition to} the
classical zero-point spectrum $U^{zp}\left(  \omega\right)  $, so that both
are present,%

\begin{equation}
U^{total}\left(  \omega,T\right)  =U^{T}\left(  \omega,T\right)
+U^{zp}\left(  \omega\right)  . \label{Utotneed}%
\end{equation}
If the radiation in the momentarily comoving inertial frame includes thermal
radiation, we expect that the total radiation spectrum will continue to
include thermal radiation in the Rindler frame.

\subsubsection{Tolman-Ehrenfest Relation\ }

Since the thermal radiation pattern is stationary in the Rindler frame, we
expect it to satisfy the Tolman-Ehrenfest relation\cite{Tolman} connecting the
temperature to the coordinates as $T\sqrt{\left\vert g_{00}\right\vert
}=const$. \ For our Rindler coordinates in flat spacetime where $ds^{2}%
=\xi^{2}d\eta^{2}-d\xi^{2}-dy^{2}-dz^{2}$, this corresponds to $T\sqrt
{\left\vert g_{00}\right\vert }$ $=\left[  T\left(  \xi\right)  \right]
\xi=const$. \ Thus the temperature $\left[  T\left(  \xi\right)  \right]
=const/\xi$ of the radiation is varying with the spatial coordinate in the
Rindler frame, being large for small values of $\xi$ near the event horizon,
and falling smoothly toward zero at large values of $\xi$ far from the event
horizon. \ 

\subsubsection{Spectrum Remains Thermal Radiation in the Comoving Inertial
Frame\ }

In the momentarily comoving inertial frame, where the time coordinate
$t^{\prime}$ agrees (through first order) with the time coordinate of the
Rindler frame, there is just one temperature $T.$ \ The spatial behavior of
the radiation shows exactly one special wavelength \ $l_{T}=const/\left(
k_{B}T\right)  $ associated with the temperature $T$. \ The preferred inertial
frame is the frame of the box, which is also that of the comoving reference
frame. \ Thus, in the comoving inertial frame, at a single time, the spectrum
can indicate the presence of thermal radiation of a single temperature $T$ and
a single preferred wavelength $l_{T}$. \ The \textit{spatial} randomness is
\textit{unchanged} from that in the momentarily comoving inertial frame.

\subsubsection{Time is the Crucial Aspect Between Coordinate Frames}

It is the \textit{time} behavior in the Rindler coordinate frame that is so
completely different from that in an \textit{inertial} frame. \ Any
\textit{inertial} frame has all its clocks synchronized, and shows one
preferred length $l_{T}=const/\left(  k_{B}T\right)  $ for thermal radiation
at a single temperature $T$. \ This single temperature $T$ is the same one
indicated by the \textit{spatial} correlations inside the accelerating box.
\ However, at a single spatial point $\left(  \xi,y,z\right)  $ in the
\textit{Rindler} frame, the correlation function for the total field is larger
than that of the zero-point radiation, but varies with the distance $\xi$ from
the event horizon. \ 

If we consider only very large frequencies $\hbar\omega$ compared to $k_{B}T$,
then the only radiation present is zero-point radiation $U^{zp}\left(
\omega\right)  $. \ A \textit{single radiation mode} in the Rindler frame is
not a normal mode in the momentarily comoving inertial frame. \ The Rindler
mode will correspond to low inertial frequency near the event horizon and
correspond to high inertial frequency at large distances from the event
horizon. This follows from the relation in Eq. (\ref{Rindl}) since, for
constant $\xi,$ we have $ct=\xi\sinh\eta$ which gives $\left(  cdt\right)
=\left(  d\eta\right)  \xi\cosh\eta$ or
\begin{equation}
\frac{d}{dt}=\frac{c}{\xi\cosh\eta}\frac{d}{d\eta}.
\end{equation}
The decrease in the temperature at large distances from the event horizon is
consistent with different frequencies of inertial radiation found in the
Rindler frame.

This surprising \textit{relativistic} aspect is something which physicists
often use these days, but are frequently unaware of. \ Thus, the GPS-locating
information on cell phones depends upon \textit{time} corrections which
recognize that lower clocks run slower in the earth's gravitational field.
\ However, the time corrections are so small, that people living in tall
apartment buildings are completely unaware that clocks on the ground floor run
slower than the clocks on the top floor. \ However, it is these small
relativistic corrections which are so crucial to understanding relativistic
thermal radiation.

\subsection{Introduction of a Conformal-Invariant Constant $\zeta$ into the
Rindler Frame}

As one goes to ever-larger values of $\xi$, the temperature of the thermal
radiation also decreases, $T\left(  \xi\right)  =const/\xi$. \ \ However, we
would like to obtain the spectrum of thermal radiation. \ Thus we will write
the temperature as
\begin{equation}
T\left(  \xi\right)  =\zeta/\xi,
\end{equation}
where $\zeta$ is some \textit{temperature-related} parameter. \ 

The multiplication by $\zeta$ does not change the allowed normal modes for the
box, the Dirichlet boundary conditions at the walls of the box, or the
distance $\xi$ to the event horizon. \ This is still a steady-state
distribution in the Rindler frame. \ However, the new constant $\zeta$ related
to time does introduce a variable real constant $\zeta$ which is at our
disposal, and which should be related to the temperature. \ Thus in equation
(\ref{corrzp}), we change the coordinate time difference $\eta^{\prime}-\eta$
over to $\zeta\left(  \eta^{\prime}-\eta\right)  =\left(  \zeta/\xi\right)
\left(  c\tau^{\prime}-c\tau\right)  .$ We have used $ds^{2}=\xi^{2}d\eta
^{2}-d\xi^{2}-dy^{2}-dz^{2}$ for fixed $\xi$ to connect the proper time
interval $\tau^{\prime}-\tau$ to the Rindler coordinate time, $cd\tau=\xi
d\eta$. \ With this new constant $\zeta$, the equation (\ref{corrzp}) becomes
\begin{align}
\left\langle \phi^{total}(ct,\xi,y,z)\phi^{total}\left(  ct^{\prime}%
,\xi,y,z\right)  \right\rangle  &  =\frac{\hbar c}{4\pi}\left(  \frac{\zeta
}{\xi}\right)  ^{2}\frac{-1}{\sinh^{2}\left[  \zeta\left(  \eta^{\prime}%
-\eta\right)  /2\right]  }\nonumber\\
&  =\frac{\hbar c}{4\pi}\left(  \frac{\zeta}{\xi}\right)  ^{2}\frac{-1}%
{\sinh^{2}\left[  \zeta\left(  c\tau^{\prime}-c\tau\right)  /\left(
2\xi\right)  \right]  } \label{Z}%
\end{align}
in the Rindler frame. \ We notice that for small separation in time,
$\tau-\tau^{\prime},$ the constant $\zeta$ disappears, and the the correlation
function goes over to the zero-point radiation result in Eq. (\ref{limit}),%
\begin{align}
\left\langle \phi^{total}(ct,\xi,y,z)\phi^{total}\left(  ct^{\prime}%
,\xi,y,z\right)  \right\rangle  &  \rightarrow\frac{\hbar c}{4\pi}\left(
\frac{\zeta}{\xi}\right)  ^{2}\frac{-1}{\left[  \zeta\left(  c\tau^{\prime
}-c\tau\right)  /\left(  2\xi\right)  \right]  ^{2}}\nonumber\\
&  =\frac{\hbar c}{\pi}\frac{-1}{\left(  c\tau^{\prime}-c\tau\right)  ^{2}}.
\end{align}
In the momentarily comoving inertial frame, the proper time $\left(
c\tau^{\prime}-c\tau\right)  $ agrees with the local time, $\left(
c\tau^{\prime}-c\tau\right)  =ct^{\prime}-ct.$

\subsubsection{Fourier Frequency Transform}

For the correlation function in Eq. (\ref{Z}), we would like to know the
associated frequency spectrum in a Minkowski inertial frame which would give
such a correlation function in a Rindler frame. \ The Fourier transformation
of the function $\omega\coth\left[  \alpha\omega\right]  $ is\cite{G-R}%

\begin{align}
&  \int_{0}^{\infty}d\omega\left(  \omega\coth\left[  \frac{\xi}{\zeta}%
\frac{\pi\omega}{c}\right]  \right)  \cos\left[  \omega t\right] \nonumber\\
&  =\int_{0}^{\infty}d\omega\,\omega\cos\left[  \omega t\right]  +\int%
_{0}^{\infty}d\omega\left(  \frac{2\omega}{\exp\left[  2\pi\xi\omega/\left(
\zeta c\right)  \right]  -1}\right)  \cos\left[  \omega t\right] \nonumber\\
&  =-\frac{1}{t^{2}}+\left[  \frac{1}{t^{2}}-\left(  \frac{\zeta c}{2\xi
}\right)  ^{2}\text{csch}^{2}\left(  \frac{\zeta c}{2\xi}t\right)  \right]
=-\left(  \frac{\zeta c}{2\xi}\right)  ^{2}\text{csch}^{2}\left(  \frac{\zeta
c}{2\xi}t\right)  , \label{Fourierw}%
\end{align}
where we have used the singular integral%
\begin{equation}
\int_{0}^{\infty}dk\,k\cos\left(  bk\right)  =\operatorname{Re}\lim
_{\lambda\rightarrow0}\int_{0}^{\infty}dk\,k\exp\left[  \left(  ib-\lambda
\right)  k\right]  =-1/b^{2}.
\end{equation}
Then writing csch$\left(  x\right)  =1/\left[  \sinh\left(  x\right)  \right]
$, our correlation function in Eq. (\ref{Z}) becomes%

\begin{align}
&  \left\langle \phi^{total}\left(  ct,\xi,y,z\right)  \phi^{total}\left(
ct^{\prime},\xi,y,z\right)  \right\rangle \nonumber\\
&  =\left(  \frac{\hbar c}{4\pi}\right)  \left(  \frac{\zeta}{\xi}\right)
^{2}\frac{-1}{\sinh^{2}\left[  \zeta\left(  ct-ct^{\prime}\right)  /\left(
2\xi\right)  \right]  }\\
&  =\left(  \frac{\hbar c}{4\pi}\right)  \left(  \frac{\zeta}{\xi}\right)
^{2}\left(  \frac{2\xi}{\zeta c}\right)  ^{2}\int_{0}^{\infty}d\omega\left(
\omega\coth\left[  \frac{\pi\xi}{\zeta c}\omega\right]  \right)  \cos\left[
\omega\left(  t-t^{\prime}\right)  \right]  .
\end{align}
We note that this correlation function depending on $\zeta$ can be separated
into two pieces as
\begin{align}
&  \left\langle \phi^{total}\left(  ct,\xi,y,z\right)  \phi^{total}\left(
ct^{\prime},\xi,y,z\right)  \right\rangle \nonumber\\
&  =\left(  \frac{-\hbar}{\pi c}\right)  \left\{  \left[  \left(  \frac{\zeta
c}{2\xi}\right)  ^{2}\text{csch}^{2}\left(  \frac{\zeta c}{2\xi}\left(
t-t^{\prime}\right)  \right)  -\frac{1}{\left(  t-t^{\prime}\right)  ^{2}%
}\right]  +\frac{1}{\left(  t-t^{\prime}\right)  ^{2}}\right\}  .
\end{align}
The part involving the square brackets is finite as $\left(  t-t^{\prime
}\right)  \rightarrow0,$ and corresponds to the thermal radiation, whereas the
part $1/\left(  t-t^{\prime}\right)  ^{2}$ diverges as $\left(  t-t^{\prime
}\right)  \rightarrow0$ and corresponds to the zero-point radiation. \ 

\subsubsection{Connection of $\zeta/\xi$ with Temperature}

Now we want to integrate over the finite \textit{thermal} part of the spectrum
at a fixed coordinate $\xi$ to obtain the connection of $\zeta/\xi$ with
temperature $T$ as given in the Stefan-Boltzmann law, $u\left(  T\right)
=a_{S}T^{4}$. \ Subtracting off the zero-point radiation spectrum so as to
leave the thermal part of the spectrum, we find%

\begin{align}
&  \int_{0}^{\infty}d\omega\frac{\omega^{2}}{\pi^{2}c^{3}}\left(  \frac{1}%
{2}\hbar\omega\coth\left[  \frac{\pi\xi}{\zeta c}\omega\right]  -\frac{1}%
{2}\hbar\omega\right) \nonumber\\
&  =\left(  \frac{\hbar c}{2\pi^{2}c^{4}}\right)  \left(  \frac{\zeta c}%
{\pi\xi}\right)  ^{4}\int_{0}^{\infty}du\,u^{3}\left(  \coth\left[  u\right]
-1\right) \nonumber\\
&  =\left(  \frac{\hbar c}{2\pi^{2}c^{4}}\right)  \left(  \frac{\zeta c}%
{\pi\xi}\right)  ^{4}\frac{1}{8}\frac{\pi^{4}}{15}=a_{S}T^{4}=\frac{\pi
^{2}k_{B}^{4}}{15\hbar^{3}c^{3}}T^{4},
\end{align}
where we have inserted the value $a_{S}=\left(  \pi^{2}k_{B}^{4}\right)
/\left(  15\hbar^{3}c^{3}\right)  .$ \ Therefore the connection with
temperature is%

\begin{equation}
T=\frac{\left(  \zeta c\right)  \hbar}{2\pi\xi k_{B}}. \label{T}%
\end{equation}
But now we can replace the expression $\zeta/\xi$ by the familiar temperature
$T$. \ Comparing with Eq. (\ref{Utotneed}), we have found exactly the Planck
spectrum including zero-point radiation%

\begin{align}
U^{total}\left(  \omega,T\right)   &  =U^{T}\left(  \omega,T\right)
+U^{zp}\left(  \omega\right)  =\frac{1}{2}\hbar\omega\coth\left[  \frac
{\hbar\omega}{2k_{B}T}\right] \nonumber\\
&  =\frac{\hbar\omega}{\exp\left[  \hbar\omega/\left(  k_{B}T\right)  \right]
-1}+\frac{1}{2}\hbar\omega.
\end{align}
Thus by considering the zero-point radiation in a Rindler frame, we are led to
the Planck spectrum including zero-point radiation in a Minkowski frame. \ The
presence of thermal radiation at a temperature above zero, is required for a
positive value of the constant $\zeta$. \ At zero temperature, $\zeta/\xi=0$
requires $\zeta=0$. \ 

\section{Summary}

In this article, we suggest criteria for zero-point radiation and for thermal
radiation within classical physics. \ Using these criteria, we give a
derivation of the Planck spectrum plus zero-point radiation for relativistic
scalar waves. \ We emphasize that relativistic waves satisfy conformal
symmetry in Minkowski spacetime. \ Classical zero-point radiation is a
spectrum of random classical radiation which is conformal \textit{invariant}.
\ On the other hand, \textit{thermal} radiation for relativistic classical
wave fields is characterized by exactly one \textit{variable} parameter, its
temperature $T$, which changes under a $\sigma_{ltU^{-1}}$-dilation of the
conformal group. \ Thus zero-point radiation and thermal radiation are quite
distinct in classical physics; the zero-point radiation is $\sigma_{ltU^{-1}}%
$-dilation \textit{in}variant\textit{,} and the thermal radiation is
\textit{co}variant, changing with a change in scale as, $T^{\prime}=T/\sigma,$
where $\sigma$ is a positive real number. \ The crucial aspect is that
\textit{in a Rindler frame}, both zero-point radiation and thermal radiation
take exactly the same \textit{functional} form with only one constant
(depending on temperature) being different. \ \ 

\section{Acknowlegement}

I wish to thank Professor Daniel C. Cole for reading the manuscript and for
pointing out ambiguous passages and typos.

March 20, 2026 \ \ \ \ \ RindlerPlanck6.tex

\end{document}